\documentclass[10pt,aas2pp4]{aastex}
\usepackage{amssymb}
\usepackage{amsmath}
\usepackage{graphicx}
\usepackage{lscape}
\usepackage{rotating,color}
\usepackage{color}

\begin{document}

\title{The growth of the disk galaxy UGC8802}



\author{R. X. Chang \altaffilmark{1}; S. Y. Shen \altaffilmark{1}; J. L. Hou \altaffilmark{1};}

\altaffiltext{1}{ Key Laboratory for Research in Galaxies and Cosmology,
  Shanghai Astronomical Observatory, Chinese Academy of Sciences, 80 Nandan Road, Shanghai, China, 200030 }
 \email{crx@shao.ac.cn}

\begin{abstract}

The disk galaxy UGC8802 has high neutral gas content and a flat profile of star
formation rate compared to other disk galaxies with similar stellar mass. It
also shows a steep metallicity gradient. We  construct a chemical evolution
model to explore its growth history by assuming its disk grows gradually from
continuous gas infall, which is shaped by a free parameter -- the
infall-peak time. By adopting the recently observed molecular surface density
related star formation law, we show that a late infall-peak time can naturally 
explain the observed high neutral gas content, while an inside-out
disk formation scenario can fairly reproduce the steep oxygen 
abundance gradient. Our results show that most of the observed 
features of UGC8802 can be well reproduced by simply `turning 
the knob' on gas inflow with one single parameter, which implies that 
the observed properties of gas-rich galaxies could also be modelled 
in a similar way. 

\end{abstract}


\section{Introduction}

Understanding the content and distribution of cold gas in galaxies is an
important step to understand the formation and evolution of galaxies. UGC8802
is an interesting target selected from the \sl GALEX \rm Arecibo SDSS Survey
(GASS) \citep{gar09,cat10}. One interesting aspect is its cold gas content.
UGC8802 is a disk galaxy with stellar mass $M_*\approx 2 \times 10^{10}
\rm{M_{\odot}}$, while its HI mass is estimated to be as high as $2.1 \times
10^{10} \rm{M_{\odot}}$ \citep{sprin05} and its molecular gas mass is only
about one-tenth of its HI mass \citep{moran10}. Compared to other disk galaxies in this
stellar mass range, the high neutral gas content of UGC8802 is uncommon
\citep{gio07,gar09,cat10}. Why UGC8802 has such high neutral gas
fraction and how UGC8802 acquires its cold gas are still open questions
\citep{moran10}.

Another interesting aspect of UGC8802 is its flat profile of the star
formation rate (SFR) \citep{moran10}. The SFR surface densities of  `normal'
disk galaxies typically decrease (sometimes exponentially) from the center to
the outer regions of the disk, such as the Milky Way, M31, etc
\citep{big08,fu09,yin09}. Besides the flat SFR profile, \citet{moran10}
found that UGC8802 shows a steeper radial oxygen gradient than the average
gradient of the GASS sample.

Therefore, it is interesting to explain these observed results under the
frame of the current models of galaxy formation and evolution. For individual
disk galaxy, parameterized modeling is proven to be fruitful in exploring
their galactic formation and evolution
\citep{tin80,chang99,bp00,chi01,col09,yin09}. In this letter, we construct
such a parameterized model to investigate the radial-dependent star formation history
(SFH) of UGC8802 and so as to explore its growth history.

\section{The model}

Our model is based on that of \citet{chang10}. Here we emphasize main
ingredients of the model, especially some revisions. We adopt the standard
cold dark matter (CDM) cosmology with $\Omega_{\rm{M}}=0.3,\Omega_{\Lambda}=0.7$
and $H_0=70 \rm {km s^{-1} Mpc^{-1}}$.

We assume that the galactic disk is sheet-like and composed by a set of
independent rings, each 500pc wide. The disk originates through continuous
gas-infall from the dark halo. We assume that the disk begins to form at 12.5
$\rm{Gyr}$ ago, which roughly corresponds to $z\sim 6$ under the
standard cosmology. At given radius $r$, the gas infall rate $f_{\rm{in}}(r,t)$ (in
units of $\rm{M_{\odot}} \rm {pc}^{-2} \rm{Gyr}^{-1}$) is assumed to be
Gaussian in time \citep{chang99,chang10}:
\begin{equation}
f_{\rm{in}}(r,t)= \frac{A(r)}{\sqrt{2\pi}\sigma}e^{-[t-t_{\rm{p}}(r)]^2/2\sigma^2}
\end{equation}
where $t_{\rm{p}}$ is the infall-peak time and $\sigma$ is the scatter. We adopt $t_{\rm{p}}$
to be free and set $\sigma=3\rm{Gyr}$ since the model results are insensitive to 
varying $\sigma$ in the $2\sim 4 \rm{Gyr}$ range \citep{chang10}. The $A(r)$ are a set of separate
quantities normalized by the stellar mass surface density of present-day
$\Sigma_*(r,t_{\rm{g}})$, where $t_{\rm{g}}$ is the cosmic age and set to be
$t_{\rm{g}}=13.5\rm{Gyr}$ according to the adopted cosmology. The stellar mass of
UGC8802 and its disk scale-length at the present time are adopted to be
$M_*=2\times 10^{10} \rm{M_{\odot}}$ and $r_{\rm{d}}=5.8\rm{kpc}$ \citep{moran10}. We
assume that the galaxy UGC8802 at present is a pure-disk system and
$\Sigma_*(r,t_{\rm{g}})$ follows an exponential profile, and so that the central
stellar mass surface density is given as $\Sigma_*(0,t_{\rm{g}})=M_*/(2\pi r_{\rm{d}}^2)$.

The star formation (SF) law is one of the key ingredients of our disk galaxy
formation model. \citet{ken98} found a power-law relationship between the
galaxy-averaged SFR surface density and the galaxy-averaged total gas surface
density, which is often called as the classical Kennicutt-Schmidt
SF law and has been widely used in the chemical evolution models of the disk
galaxy formation. Later, extended analyses to kpc-scale
regions within several Local Group galaxies show that the SFR surface density
correlates stronger with the surface density of molecular hydrogen than that
of atomic hydrogen \citep{wong02,big08,leroy08}.

In this study, we adopt a recently observed molecular surface density
related SF law (hereafter, it is termed as the $\Sigma_{\rm{H_2}}$-based SF law), that is, the SFR
surface density $\Psi(r,t)$ is linearly proportional to the molecular
hydrogen surface density $\Sigma_{\rm{H_2}}(r,t)$:
\begin{equation}
\Psi(r,t)=\Sigma_{\rm{H_2}}(r,t)/t_{\rm{dep}},
\end{equation}
where $t_{\rm{dep}}$ is the molecular gas depletion time.  \citet{big08} and
\citet{leroy08} derived a constant molecular gas depletion time as $t_{\rm{dep}}=2
\rm{Gyr}$ in their sample, while \citet{sain11} found the non-universality of
$t_{\rm{dep}}$ in galaxies in the CO Legacy Database for the GASS (COLD GASS)
sample. \citet{sain11} found that the strongest dependence of $t_{\rm{dep}}$ is on the
stellar mass and the mean $t_{\rm{dep}}$ is parameterized as
log$(t_{\rm{dep}}/\rm{yr})=0.36(log M_*/M_{\odot}-10.7)+9.3$. According to this
relation, we adopt $t_{\rm{dep}}=0.77 \rm{Gyr}$ for UGC8802.

Regarding the molecular-to-atomic mass fraction $R_{mol}(r,t)$, \citet{br06}
obtained an empirical fit relation:
\begin{equation}
R_{\rm{mol}}(r,t)=\Sigma_{\rm{H_2}}(r,t)/\Sigma_{\rm{HI}}(r,t)=(P_{\rm{h}}/P_{{\rm{h,0}}})^\gamma,
\end{equation}
where $P_{\rm{h}}$ is the mid-plane pressure, $P_{\rm{h,0}}$ and $\gamma$ are fitting constants
and we adopt $P_{\rm{h,0}}/k=1.7\times10^4\rm{cm}^{-3}\rm{K}$
and $\gamma=0.8$ \citep{leroy08}. The pressure is estimated to include gas
self-gravity  \citep{elm89,leroy08}:
\begin{equation}
P_{\rm{h}}=\frac{\pi}{2}\rm{G}\Sigma_{gas}(\Sigma_{gas}+\frac{\sigma_g}{\sigma_{*,z}}\Sigma_*),
\end{equation}
where $\Sigma_{\rm{gas}}$ is the total gas surface density, $\sigma_{\rm{g}}$ and
$\sigma_{*,z}$ are the vertical velocity dispersion of gas and stars. We
adopt $\sigma_{\rm{g}}=11\rm{km}\rm{s}^{-1}$ \citep{ostrik10}, while $\sigma_{*,z}$
is estimated as [see appendix B of \citet{leroy08} for details]:
\begin{equation}
\sigma_{*,z}=\sqrt{\frac{2\pi\rm{G}r_{\rm{d}}}{7.3}}\Sigma_*^{0.5}.
\end{equation}

A similar $\Sigma_{\rm{H_2}}$-based SF law has been successfully adopted in the
semi-analytical model of galaxy formation by \citet{fu10}. \citet{kang12} also use
the $\Sigma_{\rm{H_2}}$-based SF law to explore the chemical evolution and
SFH of M33 and find that, comparing to the model adopting the classical Kennicutt-Schmidt
SF law, the model adopting the $\Sigma_{\rm{H_2}}$-based SF law predicts steeper color and
matellicity gradients, which are more consistent with the observations.

Other components of the model, such as the initial mass faction (IMF),
chemical evolution, etc., are the same as that of \citet{chang10}. We
also consider the contribution of gas outflow process by assuming that the
gas outflow rate is proportional to the SFR and the coefficient is set to be
$b_{\rm{out}}=0.007$ according to the mass-dependent model of
\citet{chang10}. We point out that our final results are insensitive to the
variation of $b_{\rm{out}}$ since UGC8802 is a massive galaxy and the gas
outflow process is not a dominant process of its chemical evolution. 

We emphasize that our model only has one free parameter, the
infall-peak time $t_{\rm{p}}$. In \citet{chang10}, $t_p$ is assumed to 
be a function of galactic stellar mass and do not vary with radius. 
In the case of UGC8802, we assume that $t_{\rm{p}}$ is a function of 
radius since we will further consider the radial profiles of the UGC8802 
rather than the global properties of galaxies as that in \citet{chang10}. 
Since the formalizing of $t_{\rm{p}}(r)$ regulates the shape of SFHs 
along the disk, we aim to get the constraints on $t_{\rm{p}}(r)$ from 
the observed properties of UGC8802 and then get insights on its
growth history.

\section{Observations versus model predictions}

We present our results step by step. Firstly, we summarize main observational
properties of UGC8802, and then explore the influence of the free parameter
on our results. Finally, we present a viable model to discuss the 
radial-dependent SFH of UGC8802.

The observed data of the radial profiles of UGC8802 are adopted mainly from
\citet{moran10} and plotted in Fig. 1. Since our model only has the
galactic-center distance and cannot distinguish the negative position from
the positive one, we plot the observed data of the negative and positive
positions as open triangles and filled squares in Fig. 1, respectively. The
NUV-r color profile of UGC8802 is measured using SDSS and GALEX photometry
\citep{moran10}, but the data has not been corrected for dust-attenuation.
The observed data of oxygen abundance $\rm{log}(O/H)+12$, the SFR surface
density and $D_n(4000)$ are obtained from the long-slit spectroscopy of
UGC8802 \citep{moran10}. The data $SFR/<SFR>$ are estimated from the observed
SFRs and the stellar surface density $\Sigma_*(r,t_{\rm{g}})$ by assuming $(1-R)$
fraction of the total formed stellar mass are locked in the stellar mass in
the present day, where $R$ is the return fraction and we set $R=0.3$ according
to the adopted IMF.

The comparison between model predictions of the radial profiles and the
observations are presented in Fig.1. The dash and dash-dot lines are model
results of two limiting cases of $t_{\rm{p}}=0.1\rm{Gyr}$ and $t_{\rm{p}}=15\rm{Gyr}$,
respectively. The case $t_{\rm{p}}=0.1\rm{Gyr}$ (dash lines) corresponds to a
time-decreasing gas-infall process that most of the gas has been accreted in
the early stage of its history, while that of $t_{\rm{p}}=15\rm{Gyr}$ (dash-dot
lines) represents a time-increasing gas accretion that there is still a large
fraction of cold gas-infall at the present time. Fig. 1 shows that the model
predictions are very sensitive to the adopted $t_{\rm{p}}$. The model adopting the
later infall-peak time (dash-dot lines) predicts the higher surface density
of $\rm{H_2}$ and $\rm{HI}$, the higher SFR and specific SFR, the lower
gas-phase metallicity, the older mean stellar age and the bluer color. These
results are straightforward, the later $t_{\rm{p}}$ corresponds to a more recent gas
infall and then results in a younger stellar population and a higher fraction
of cold gas in present.

\begin{figure}[!t]
  \centering
  \includegraphics[height=8cm,width=8cm]{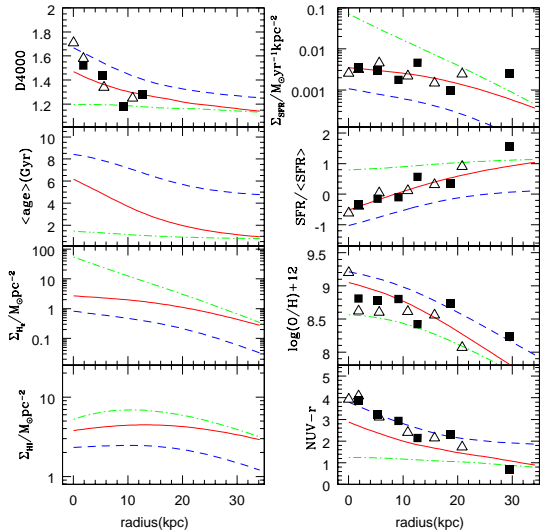}\\
\caption{Comparisons of the radial profiles between the observations and the model predictions.
The dash and dash-dot lines are model results adopting $t_{\rm{p}}=0.1\rm{Gyr}$ and $t_{\rm{p}}=15\rm{Gyr}$,
respectively. The solid lines plot the predictions of the viable model,
which adopts $t_{\rm{p}}(r)/{\rm{Gyr}}=1.5r/r_{\rm{d}}+5$.
The observed data are described in details in the text. }
\end{figure}

Fig. 1 also shows that the area between the dash and dash-dot lines almost
covers the locations of the observed data. There is a trend that the observed
data at the inner region approaches the dash lines (the early $t_{\rm{p}}$) and the
data at the  outer regions approaches the dash-dot lines (the late $t_{\rm{p}}$).
After a set of calculations and comparisons, we select a viable model by
adopting an inside-out formation scenario $t_{\rm{p}}(r)/{\rm{Gyr}}=1.5r/r_{\rm{d}}+5.0$ and
plot its model results as solid lines in Fig. 1. This model predicts
$M_{\rm{H_2}}=3.7 \times 10^{9} \rm{M_{\odot}}$ and $M_{\rm{HI}}=1.5 \times 10^{10}
\rm{M_{\odot}}$, which in good agreement with the observed
values [$M_{\rm{HI}}=2.1 \times 10^{10} \rm{M_{\odot}}$, $M_{\rm{H_2}}=1.45
\times 10^{9} \rm{M_{\odot}}$ \citep{moran10}], considering the observed
uncertainties. Our model-predicted $NUV-r$ colors are systematically bluer
than the observed ones, which is also reasonable since the observed $NUV-r$
color has not been corrected for dust-attenuation and our model predictions  
are dust free.

\begin{figure}[!t]
  \centering
  \includegraphics[height=7cm,width=8cm]{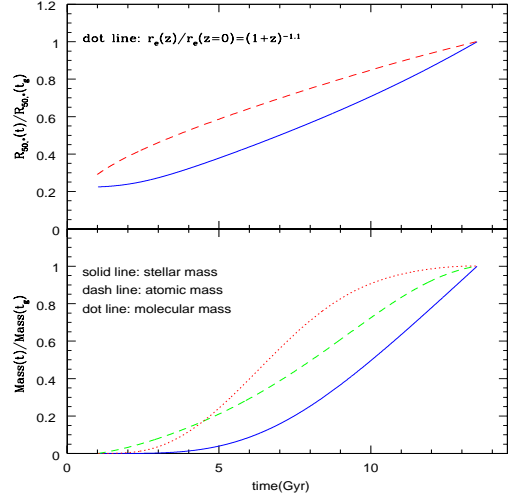}\\
\caption{The growth history of UGC8802 predicted by the viable model. 
In the upper panel, the solid line shows the model predicted time evolution of the half mass
size $R_{50,*}$, while the dash line shows the observed trend of size
evolution of massive galaxies. The bottom panel shows the normalized evolution history of
different baryonic components, including the stellar mass (the solid line), the atomic hydrogen
mass $M_{\rm{HI}}$ (the dash line) and the molecular hydrogen mass $M_{\rm{H_2}}$ (the
dot line).}
\end{figure}

The good  consistence between our model predictions and  the observations
implies that the continue gas-infall model is also viable for the `peculiar'
UGC8802, which has steeper metallicity gradient, flatter SFR profile and
extraordinary higher neutral gas fraction than normal disk galaxies. The key
ingredient of this successful model is that the infall-peak time $t_{\rm{p}}(r)$
increases from $5\rm{Gyr}$ in the innermost disk to 12.5Gyr at the out region
around $\sim 5 r_{\rm{d}}$. That means the formation of UGC8802 is quite late and
the disk growing is still ongoing. In fact, this scenario is the
well-known idea that the disk forms inside-out and has also been applied in
previous models of formation and evolution of disk galaxies
\citep{chang99,bp00,chi01,col09,fu09,yin09,wang10}.

The above results have shown that the inside-out growth scenario can
nicely reproduce the main observed properties of UGC8802. In Fig. 2, we show
further predictions of the growth history of the UGC8802 of this established
model. The upper panel of Fig.2 shows the model predicted time evolution of
the half-size stellar mass $R_{50,*}$ (solid line), which is defined as the
radius where contains half of the total stellar mass. As we can see, our
model predicts that the half-size of UGC8802 increases almost linearly with
time, which is also globally consistent with the observations of the average size evolution
trend of the massive 
galaxies [$r_{\rm{e}}(z)/r_{\rm{e}}(z=0)=(1+z)^{-1.1}$, \citep{van08}, dash line]. 
This size growth is actually another angle of view of the disk inside-out formation
scenario.

In the bottom panel, we plot the model predictions of the evolution of 
the different mass components, which includes the stellar mass $M_*$ (solid
line), the atomic hydrogen mass $M_{\rm{HI}}$ (dash line) and the molecular
hydrogen mass $M_{\rm{H_2}}$ (dot line). The quantities are
normalized by its value at the present-day.  The bottom panel of Fig. 2 shows
that, in the early epoch, with the continue infall of cold gas (mainly in the
form of the atomic gas), the molecular gas accumulates faster, while the
accumulation of stellar mass is lagged behind. Later, the mass of the
molecular gas becomes to be asymptotic, while the growth rate of stellar mass
is speeded. Indeed, half of the stellar mass has been accumulated during the
last $4 \rm{Gyr}$. To sum up, our results suggest that UGC8802 is a young
galaxy and still active in star formation processes.

\section{Summary}

In this work, we build a bridge for disk galaxy UGC8802 between its observed
properties and its growth history by constructing a simple chemical
evolution model. We find that a late infall-peak time $t_{\rm{p}}$, which
corresponds to a recent gas-infall process, is necessary to explain the
observed high atomic and molecular gas content of UGC8802. The model adopting
an inside-out formation scenario $t_{\rm{p}}(r)/{\rm{Gyr}}=1.5r/r_{\rm{d}}+5.0$ and the
$\Sigma_{\rm{H_2}}$-based SF law can reproduce the observed radial profiles
of UGC8802 very well. This consistency
does encourage us that our simple model is viable for such a `peculiar'
galaxy. This result implies that there should be no violent star burst
process (e.g. the major merging) during the growth history of UGC8802.
According to our model, half of the stellar mass of UGC8802 may have
been accumulated during the last $4 \rm{Gyr}$ and the SF process of the
outer disk is still active. The physics behind the very young history 
of UGC8802 may related to its environment, however, 
which is beyond the scope of this letter.

The success of our simple model on the case of UGC8802 implies
that the observed properties of the gas-rich galaxies might also be well
modeled by `tuning the knob' on gas inflow with a single parameter.
Indeed, \citet{wang10} and \citet{moran11} find that the color and metallicty of
the outer region of the galaxies strongly correlate with its total atomic mass $M_{\rm{HI}}$, i.e.,
the higher the $M_{\rm{HI}}$, the bluer the color, the lower the metallicity. In our
story, the high atomic gas fraction of gas-rich galaxy may be the result 
of a late and smooth gas accretion history, and the bluer color and lower 
metallicity of its outer region are embedded in the disk inside-out formation scenario. 
We will further explore these questions in a forthcoming paper. 

\begin{acknowledgements}

We thank the anonymous referee for suggestions to greatly improve this paper.
This work is supported by the National Science Foundation of China
No.10573028, 10403008, the Key Project No.10833005 and No. 10878003,
the Group Innovation Project No.10821302.

\end{acknowledgements}

\end{document}